\documentclass[aps,letterpaper,nofootinbib,preprint,showpacs,preprintnumbers,amsmath,amssymb]{revtex4}%
\usepackage{latexsym}%
\usepackage{epsf}%
\usepackage[hypertex]{hyperref}%

\def\oP{{\overline P}}

\def\cF{{\mathcal F}}

\def\cN{{\mathcal N}}

\def\cQ{{\mathcal Q}}
\def\cW{{\mathcal W}}

\newcommand{\beq}{\begin{equation}}
\newcommand{\beqn}{\begin{equation}\nonumber}
\newcommand{\eeq}{\end{equation}}
\newcommand{\bea}{\begin{eqnarray}}
\newcommand{\bean}{\begin{eqnarray}\nonumber}
\newcommand{\eea}{\end{eqnarray}}

\begin{document}

\begin{center}
{\bf{\Large Canonical Quantization of Spherically Symmetric Dust Collapse}}
\bigskip
\bigskip

{{Cenalo Vaz$^{a,b,}$\footnote{e-mail address: cenalo.vaz@uc.edu},
Louis Witten$^{b,}$\footnote{e-mail address: lwittenw@gmail.com}}}
\bigskip

{\it$^a$RWC and $^b$Department of Physics,}\\
{\it University of Cincinnati,}\\
{\it Cincinnati, Ohio 45221-0011, USA}
\end{center}
\bigskip
\bigskip
\medskip

\centerline{ABSTRACT}
\bigskip\bigskip
Quantum gravity effects are likely to play a crucial role in determining the outcome of 
gravitational collapse during its final stages. In this contribution we will outline a 
canonical quantization of the LeMaitre-Tolman-Bondi models, which describe the collapse of 
spherical, inhomogeneous, non-rotating dust. Although there are many models of gravitational 
collapse, this particular class of models stands out for its simplicity and the fact that 
both black holes and naked singularity end states may be realized on the classical level, 
depending on the initial conditions. We will obtain the appropriate Wheeler-DeWitt equation 
and then solve it exactly, after regularization on a spatial lattice. The solutions describe 
Hawking radiation and provide an elegant microcanonical description of black hole entropy, 
but they raise other questions, most importantly concerning the nature of gravity's fundamental 
degrees of freedom.
\vfill\eject

\section{Introduction}

It is a great pleasure to have been invited to contribute to a festschrift in honor of Joshua N. 
Goldberg. His gentle manner, his fine intelligence and his modest nature are recognized by all. 
They conceal his ardor in defending the moral ideals he possesses and loves.

So long as no generally agreed upon theory of quantum gravity exists, it is important to examine
the quantization of particular models. This contribution addresses the quantization of the 
LeMa\^\i tre-Tolman-Bondi (LTB) solutions, which describe the classical collapse of spherically 
symmetric, inhomogeneous, self-gravitating dust. The classical solutions were originally 
introduced by G. LeMa\^\i tre \cite{l33} to study  cosmology, where it has found interesting 
applications \cite{k97}. Our principal interest here will be to develop the Hamiltonian formalism for 
both the classical and quantum LTB models, as such we will describe a generalization of work by 
Kucha\v r \cite{k95}, who developed a midisuperspace quantization of the Schwarzschild black hole. 
While it would be preferable to take a fundamental field (e.g. a scalar 
field) for the matter part, this would make the formalism much less tractable \cite{r95}. Moreover, 
the relevant features of gravitational collapse already exhibit themselves for the dust model in 
the sense that the dust collapse may result classically in the formation of a black hole or of a 
naked singularity, depending on the initial conditions.

In section II we review classical LTB collapse and present the canonical formalism for these models. 
The hypersurface action yields two constraints, {\it viz.,} the Hamiltonian constraint and the momentum 
constraint.  We reconstruct the mass and time from the canonical data and this leads naturally to new 
variables {\it viz.,} the mass function, the dust proper time, the physical radius and their conjugate momenta, 
which are introduced along with the generator of the canonical transformation from the old to the new 
variables. The momentum conjugate to the mass function may be eliminated in the Hamiltonian constraint 
using the momentum constraint. This leads to a new and simpler constraint that is able to take the place 
of the original Hamiltonian constraint. We apply Dirac's quantization program to the new constraints in 
section III and obtain the Wheeler-DeWitt equation. We then introduce a lattice regularization of the 
functional equations and find exact solutions, which we use in section IV to describe Hawking radiation. 
In section V we show how black hole entropy can be recovered from a microcanonical ensemble of states 
and discuss some issues raised by our approach in section VI.

\section{The Classical LTB Models}

The LTB models describe self-gravitating dust. The energy-momentum tensor reads $T_{\mu \nu} = 
\varepsilon(\tau,\rho) U_{\mu} U_{\nu}$, where $U^{\mu}=U^{\mu}(\tau, \rho)$ is the four-velocity vector
of a dust particle with proper time $\tau$ and labeled by $\rho$ ($\rho$ thus labels the various shells 
that together form the dust cloud). The LTB line element is given by
\beq
ds^2 = d\tau^2 - \frac{(\partial_{\rho}R)^2}{1+2E(\rho)} d\rho^2 - R^2(\rho)d\Omega^2.
\label{ltb-metric}
\eeq
Inserting this expression into the Einstein equations yields
\beq
\varepsilon(\tau,\rho) = \frac{\partial_{\rho}F}{R^2 \partial_{\rho}R} \quad \mathrm{and} \quad 
(\partial_{\tau}R)^2 = \frac{F}{R} + 2E,
\label{ltb-eg}
\eeq
where $F(\rho)$ and $E(\rho)$ are non-negative functions of the label coordinate and we have set $8\pi G=
1$. The case of collapse is described by $\partial_{\tau}R(\tau,\rho)<0$. There still exists the freedom 
to rescale the shell index $\rho$, which we fix by demanding $R(0,\rho) = \rho$, so that for $\tau=0$ 
the label coordinate $\rho$ is equal to the curvature radius $R$. Now we can express the functions $F(\rho)$ 
and $E(\rho)$ in terms of the energy density $\varepsilon(\tau,\rho)$ at $\tau=0$. From \eqref{ltb-eg},
\bea 
F(\rho) &=& \int_0^{\rho} \varepsilon(0,\tilde \rho) \tilde{\rho}^2\,d\tilde \rho,\cr
E(\rho) &=&\frac{1}{2} [\partial_{\tau}R(\tau=0,\rho)]^2-\frac{1}{\rho} \int_0^\rho
\varepsilon (0,\tilde \rho) \tilde{\rho}^2 \, d\tilde \rho,
\label{ltb_F}
\eea
so that $F(\rho)/2$ may be interpreted as the active gravitating mass inside of the shell labeled by $\rho$ 
and $E(\rho)$ as the total energy inside the shell. An analysis of the classical solutions for these models 
can be found in \cite{js95}. In the present work we discuss the canonical formalism for the so-called ``marginally 
bound'' models, defined by $E(\rho)=0$ \cite{vws01}. A generalization to non-marginally bound models is found 
in \cite{kmv06}.

\subsection{Hamiltonian formalism}

Begin with the general ansatz for a spherically-symmetric line element,
\beq
ds^2 = -N^2 dt^2 + L^2 \left( dr + N^r dt \right)^2 +   R^2 d \Omega^2,
\label{spheric_metric}
\eeq
where $N$ and $N^r$ are the lapse and shift function, respectively. The canonical momenta are given by
\bea
P_L &=& \frac{R}{N} \left(- \dot{R} + N^r R' \right),\cr
P_R &=& \frac{1}{N} \left[ -L \dot{R} - \dot{L}R + \left(N^r L R \right)'\right],
\label{P_LR}
\eea
where a dot denotes a derivative with respect to coordinate time $t$, while a prime denotes a derivative
with respect to $r$. All variables are functions of $t$ and $r$. A Legendre transformation from the 
Einstein--Hilbert action then leads to
\beq
S_{\mathrm{EH}} = \int \mathrm{d}t \int_0^{\infty} \mathrm{d}r \left(P_L \dot{L} + P_R \dot{R} - N
    H^g - N^r H_r^g \right) + S_{\partial \Sigma},
\label{mega_action}
\eeq
in which the Hamiltonian and the diffeomorphism (momentum) constraint are given by
\bea
H^g &=& - G\left(\frac{P_L P_R}{R} - \frac{LP_L^2}{2 R^2}\right) + \frac{1}{G}\left[ -\frac{L}{2} - 
\frac{R'^2}{2L}+ \left(\frac{RR'}{L} \right)'\right],\cr
H_r^g &=& R' P_R - LP_L',
\eea
respectively, and the boundary action $S_{\partial \Sigma}$ is discussed below.

The total action is the sum of \eqref{mega_action} and an action $S^d$ describing the dust. The canonical 
formalism for the latter was developed in \cite{bk95}, (see also \cite{vws01}). The dust action reads
\beq
S^d = \int \mathrm{d}t \int_0^{\infty} \mathrm{d}r \left( P_{\tau} \dot{\tau} - N H^d -
    N^r H_r^d \right),
\eeq
where
\beq
H^d = P_{\tau} \sqrt{1+\frac{{\tau'}^2}{L^2}} \quad \mathrm{and} \quad H_r^d = \tau'
P_{\tau}
\label{H_dust}
\eeq
are the dust Hamiltonian and momentum constraints respectively.

\subsection{Mass function in terms of canonical variables}

In the following we shall write the mass function $F(\rho)$, which was introduced in \eqref{ltb_F}, 
in terms of the canonical data. This is essential for deriving consistent falloff conditions that are 
appropriate for a realistic collapse model. We begin by requiring the spacetime described by the metric 
\eqref{spheric_metric} to be embedded in a LTB spacetime. Considering the LTB metric \eqref{ltb-metric}, 
a foliation described by functions $\tau(r,t)$ and $\rho(r,t)$ leads to
\bea
&&L^2 = R'^2 \rho'^2 -\tau'^2,\cr
&&N^r = \frac{R'^2 \dot \rho\rho'-\dot \tau\tau'}{L^2},\cr
&&N = \frac{R'}{L}(\dot \tau\rho'-\dot \rho\tau').
\label{lapseshift}
\eea
When these expressions for lapse function and shift vector into the expression for the canonical momentum 
$P_L$ in \eqref{P_LR} and the equations \eqref{ltb-eg} are used, we find
\beq
\frac{LP_L}{R} \stackrel{\makebox[0cm]{\scriptsize \eqref{ltb-eg}}}{=}
  R'\sqrt{1-\mathcal{F}} - \mathcal{F}\tau',
\eeq
where $\mathcal{F}\equiv 1-F/R$. Solving for $\tau'$ gives
\beq
\tau' = +\frac{1}{\mathcal{F}} \left(R' \sqrt{1-\mathcal{F}} - \frac{LP_L}R\right)
\label{lost_tau}
\eeq
and inserting this expression into \eqref{lapseshift} yields
\beq
L^2 = \left[\frac{R'^2}{\cF}- \frac{L^2 P_L^2}{R^2 \cF}\right],
\eeq
which determines $\cF$ according to
\beq
\cF = \left[\frac{R'^2}{L^2}- \frac{P_L^2}{R^2}\right].
\eeq
We can thus express $F$ locally in terms of the canonical data
as follows:
\beq
F = R \left[ 1 + \frac{P_L^2}{R^2} - \frac{{R'}^2}{L^2} \right].
\label{lost_F} 
\eeq
This expression, although obtained here for marginal models, possesses a wider range of applicability, 
holding, in fact, for {\em all} cases \cite{kmv06}. Further, it turns out that the functions, $P_F$, 
defined by
\beq
P_F = \frac{LP_L}{2R\cF}
\eeq
and the mass function, $F$, form a conjugate pair of variables. Since $R=F$ at the horizon, $\mathcal{F}=0$ 
there. We can check that though $\mathcal{F}$ appears in the denominator of \eqref{lost_tau}, $\tau'$ is 
well behaved at the horizon:
\beq
\tau' \stackrel{\mathcal{F}\rightarrow 0}{\longrightarrow} \frac{1}{2} \left(R' +
  L \right).
\eeq
This is as it should be. We now make a canonical transformation in order to elevate the mass function 
$F$ to a canonical coordinate. The canonical transformation, $(\tau, R, L, P_{\tau}, P_R, P_L) \longrightarrow 
(\tau, R, F, P_{\tau}, \bar{P}_R, P_F)$, is generated by
\beq
\mathcal G = \int_0^\infty dr \left[LP_L -\frac 12 RR' \ln \left|\frac{RR'+LP_L}{RR'-LP_L}\right|~\right]
\eeq
and this gives
\beq
\bar{P}_R = P_R - \frac{LP_L}{2R}-\frac{LP_L}{2R\mathcal{F}} - \frac{\Delta}{RL^2\mathcal{F}}, 
\eeq
with
\beq
\Delta = (RR')(LP_L)'-(RR')'(LP_L).
\eeq
The action in the new canonical variables then reads 
\beq
S_{\mathrm{EH}} = \int \mathrm{d}t \int_0^{\infty} dr\left(P_{\tau} \dot{\tau} + \bar{P}_R \dot{R} + 
P_F \dot{F} - N H - N^r H_r \right) + S_{\partial \Sigma},
\eeq
where the new constraints are
\bea
H &=& - \frac{1}{2L} \left(\frac{F'R'}{G \,\mathcal{F}} + 4 G \mathcal{F} P_F \bar{P}_R \right) + P_{\tau}
  \sqrt{1+\frac{{\tau}'^2}{L^2}},\cr
H_r &=& \tau' P_{\tau} + R' \bar{P}_R +F' P_F. 
\label{H_const_new} 
\eea
We shall now discuss the boundary action $S_{\partial \Sigma}$ in more detail.

\subsection{Boundary action}

Boundary terms are obtained from a careful discussion of the falloff conditions for the canonical 
variables, which were investigated in detail in \cite{vws01}. It turns out that the only boundary 
term is obtained from the variation of the hypersurface action with respect to $L$ and reads
\beq
\int{\mathrm d}t \ N_+(t) \delta M_+(t),
\eeq
where $N_+(t) \equiv N(t,r\to\infty)$ is the lapse function at infinity and $M_+(t)\equiv F(r\to\infty)/2$ 
is the ADM mass. To avoid the conclusion that $N_+(t)$ is constrained to vanish, which would freeze 
the evolution at infinity, the boundary term has to be canceled by an appropriate boundary action.  
This can be achieved by adding the surface action
\beq
S_{\partial \Sigma} = - \int dt  N_+(t) M_+(t).
\eeq
Since varying $N_+$ would lead to a zero ADM mass, Kucha\v{r} has argued in \cite{k95} that $N_+$ has to be 
treated as a prescribed function.  The lapse function gives the ratio of proper time to coordinate time in 
the direction normal to the foliation. Since $N^r(r)$ vanishes for $r \rightarrow \infty$, the time 
evolution at infinity is generated along the world lines of observers with $r=\mathrm{const}$. If we 
introduce the proper time, $\bar{\tau}_+$, of these observers as a new variable, we can express the lapse
function in the form $N_+(t) = \dot{\bar{\tau}}_+(t)$. This leads to
\beq
S_{\partial \Sigma} = - \int \mathrm{d}t \, M_+ \dot{\bar{\tau}}_+ . 
\label{tau_bar_bound}
\eeq
and thus we have removed the necessity of fixing the lapse function at infinity. (In \cite{k95} this is 
called `parametrization at infinities'.)

Out aim is to cast the homogeneous part of the action into Liouville form and to find a transformation to
new canonical variables that absorb the boundary terms. This can be done by introducing the mass density 
$\Gamma \equiv F'$ as a new canonical variable and using the boundary condition $F(0)=0$ (which is appropriate 
for a collapse situation). Part of the  Liouville form can then be rewritten as follows:
\bea
\bar{\theta} &\equiv & \int_0^{\infty} \mathrm{d}r \, P_{F} \delta F - M_+ \delta \bar{\tau}_+ \cr\cr
&=& \int_0^{\infty} dr   \delta \Gamma \left( \frac{\bar{\tau}_+}{2} + \int_r^{\infty} \mathrm{d}r' P_F(r') 
\right) -   \delta (M_+ \bar{\tau}_+) 
\label{new_liouville}
\eea
(see \cite{kmv06} for details). From \eqref{new_liouville} we see that $P_{\Gamma}=\bar{\tau}_+/2 + 
\int_r^{\infty} \mathrm{d}r P_F$. Thus $P_{\Gamma}(\infty)=\bar{\tau}_+/2$.  Thus the new action reads
\beq
S_{\mathrm{EH}} = \int \mathrm{d}t \int_0^{\infty} dr \left(P_{\tau}\dot{\tau} + \bar{P}_R \dot{R} + 
P_{\Gamma} \dot{\Gamma} - N H^g - N^r H_r^g \right).
\label{EH_action}
\eeq
The constraints in the new variables are
\bea
H &=& - \frac{1}{2L} \left(\frac{\Gamma R'}{G\mathcal{F}} - 4G \mathcal{F}P_{\Gamma}'
    \bar{P}_R \right) + P_{\tau} \sqrt{1+\frac{{\tau}'^2}{L^2}},\cr
H_r &=& \tau' P_{\tau} + R' \bar{P}_R
  -\Gamma P_{\Gamma}'.
\label{H1}
\eea
For the Schwarzschild black hole, $2P_{\Gamma}$ is equal to the Killing time, $T$, in the exterior, as shown 
in \cite{k95}. The Hamiltonian constraint can be greatly simplified if the momentum constraint is used to 
eliminate $P_F\equiv -P_{\Gamma}'$ (see Appendix A of \cite{kmv06}). The constraints \eqref{H1} can 
then be replaced by the following equivalent set,
\beq
H = \left({P_{\tau}}^2 + \mathcal{F} \bar{P}_R^2 \right)-\frac{\Gamma^2}{4\mathcal{F}} 
\approx 0,
\label{H_new} 
\eeq
\beq
H_r = \tau' P_{\tau} + R' \bar{P}_R - \Gamma {P_{\Gamma}}' \approx 0.
\eeq
These equations will be used as the starting point for the quantization in Sec. III. Another 
useful relation follows from \eqref{lost_tau},
\beq
\tau = T \pm \int dR \frac{\sqrt{1-\cF}}{\cF}
\eeq
and relates the dust proper time, $\tau$, to the Killing time, $T$, in the exterior of the collapsing 
dust ball, where the mass function is constant. The positive sign refers to contracting clouds and the 
negative sign to expanding clouds.

\subsection{Hamiltonian equations of motion}

Here we shall give the Hamilton equations of motion and derive Einstein's equation \eqref{ltb-eg} from them.
The Hamiltonian equations are generally given by
\bea
\dot{X} &=& \{ X, \mathcal{H}[N] + \mathcal{H}_r[N^r] \},\cr 
\dot{P}_X &=& \{P_X,\mathcal{H}[N] + \mathcal{H}_r[N^r] \},
\eea
where we have introduced the smeared constraints
\beq
\mathcal{H}[N] = \int_0^{\infty} dr N(r)H(r),~~ \mathcal{H}_r[N^r] = \int_0^{\infty} dr N^r(r)
H_r(r).
\label{super-ham-pb} 
\eeq
Starting from the action \eqref{EH_action}, the Hamiltonian equations of motion are\footnote{Note
  that $\delta F(r) / \delta \Gamma(\bar{r}) = \theta(r-\bar{r})$.}
\bea
\dot{\tau} &=&  2 N P_{\tau} + N^r \tau' \label{dottau},\cr
\dot{P}_{\tau} &=& (N_r P_{\tau})',\cr
\dot{R} &=& 2 N \mathcal{F} \bar{P}_R + N_r R',\cr
\dot{\bar{P}}_R &=&  - N \left( \frac{F \bar{P}_R^2}{R^2} + \frac{\Gamma^2 F }{4 \mathcal{F}^2 R^2} 
\right) + (N^r \bar{P}_R)',\cr
\dot{\Gamma} &=& (N^r \Gamma)' \label{dot_gamma},\cr
\dot{P}_{\Gamma} &=& N \frac{\Gamma}{2 \mathcal{F}} + N^r P_{\Gamma}' + \int_r^{\infty} d\tilde{r} 
N(\tilde{r}) \left(\frac{\bar{P}_R^2(\tilde{r})}{R(\tilde{r})} + \frac{\Gamma^2(\tilde{r})}{4
  \mathcal{F}^2(\tilde{r}) R(\tilde{r})} \right).
\eea
Consider now the momentum constraint in the form
\beq
\tau'+\frac{R'\oP_R}{P_\tau}-\frac{\Gamma P_\Gamma'}{P_\tau}=0
\eeq
and use the Hamiltonian constraint
\beq
\oP_R = \pm \frac{P_\tau}{\cF}\sqrt{\frac{\Gamma^2}{4\cF}-\cF}
\eeq
so that
\beq
\tau'=\mp \frac{R'}{\cF}\sqrt{\frac{\Gamma^2}{4\cF}-\cF}+\frac{\Gamma P_\Gamma'}
{P_\tau}=0.
\label{lost_tau2}
\eeq
Now using
\beq
P_\Gamma' = -P_F = -\frac{L P_L}{2R\cF},
\eeq
\eqref{lost_tau2} can be compared to \eqref{lost_tau} and shows that
\beq
P_\tau = \frac{\Gamma}2
\eeq
so that
\beq
\partial_{\tau} R = \frac{\dot{R}}{\dot{\tau}} \stackrel{N^r=0}{=} \frac{ 2N\mathcal{F} \bar{P}_R}
{2NP_{\tau}} = \frac{2\mathcal{F}\bar{P}_R}{\Gamma},
\eeq
which we can solve for $\bar{P}_R$. Inserting this expression into the Hamiltonian constraint 
\eqref{H_new} gives
\beq
0 = H = \frac{\Gamma^2}4 + \frac{\Gamma^2 (\partial_{\tau}R)^2}{4\mathcal{F}} - \frac{\Gamma^2}{4
\mathcal{F}}
\eeq
and solving for $(\partial_{\tau}R)^2$ leads to Einstein's equation \eqref{ltb-eg}. Note that we did not 
have to specify the lapse function.

The algebra of the constraints is not of the standard form (given, for example, in 
\cite{OUP}), because we have used the momentum constraint to eliminate $P_F$ in the Hamiltonian constraint.
In fact, a short calculation gives
\begin{align} \left\{\mathcal{H}[N], \, \mathcal{H}[M]\right\} &= 0 \; , \label{ham_poiss}\\
  \left\{\mathcal{H}_r[N^r], \, \mathcal{H}[N]\right\} &= \mathcal{H}[N_{,r}
  N^r - N N_{,r}^r] \; , \label{ham_1}\\ \left\{\mathcal{H}_r[N^r], \,
    \mathcal{H}_r[M^r]\right\} &= \mathcal{H}_r\left[[N^r,M^r]\right] \; .\label{ham_2}
\end{align}
The Poisson bracket of the Hamiltonian with itself vanishes, in contrast with the general case in which
it closes on the momentum constraint. The other brackets coincide with the general case. The transformations 
generated by the Hamiltonian constraint can thus no longer be interpreted as hypersurface deformations. 
They are in general not orthogonal to the hypersurfaces, but act along the flow lines of dust.

\section{Quantization}

We shall now apply the quantization procedure proposed by Dirac and turn the classical constraints into 
quantum operators, cf. \cite{OUP}. We begin with the expressions in \eqref{H_new}. Poisson brackets 
are translated into commutators in the Schr\"odinger representation by substituting
\beq
P_{\tau}(r) \rightarrow \frac{\hbar}{i} \,\frac{\delta}{\delta \tau(r)}, \quad
\bar{P}_{R}(r) \rightarrow \frac{\hbar}{i} \,\frac{\delta}{\delta R(r)},
\quad P_{\Gamma}(r) \rightarrow \frac{\hbar}{i} \,\frac{\delta}{\delta
  \Gamma(r)}  
\label{prescription}
\eeq 
and having them act on wave functionals. The Hamiltonian constraint \eqref{H_new} then leads to the
WDW equation,
\bea
\left[- \hbar^2 \left(\frac{\delta^2}{\delta \tau(r)^2} \right.\right. && +~ \mathcal{F} \frac{\delta^2}
{\delta R(r)^2} + A(R,F)\delta(0) \frac{\delta}{\delta R(r)} \cr 
&&\left.\left. +~ B(R,F) \delta(0)^2 \right) - \frac{\Gamma^2}{4\mathcal{F}} \right] \Psi\left[\tau,R,
\Gamma\right] = 0,
\label{WDW_eq}
\eea
where $A$ and $B$ are smooth functions of $R$ and $F$ that encapsulate the factor ordering ambiguities. 
We have introduced $\delta(0)$ in order to indicate that the factor ordering problem is unsolved and can 
be dealt with only after some suitable regularization has been performed, cf. \cite{tw87}. That is, we 
choose the terms proportional to $\delta(0)$ in such a way that the constraint algebra closes, which is 
usually called `Dirac consistency'. Quantizing the momentum constraint in \eqref{H_new} by using 
\eqref{prescription} gives
\beq
\left[\tau' \frac{\delta}{\delta \tau(r)}  + R' \frac{\delta}{\delta R(r)} - \Gamma \left(\frac{\delta}
{\delta \Gamma(r)}\right)' \right]\Psi\left[\tau, R, \Gamma \right] = 0.
\eeq
The next subsection is devoted to the application of a lattice regularization.

\subsection{Lattice regularization}

For solutions of the constraints, we make the ansatz
\beq
\Psi[\tau,R,\Gamma] = \Psi^{(0)}[F] \exp\left[-\frac i 2 \int dr~ \Gamma(r)~ \cW(\tau,R,F)\right],
\label{ansatz}
\eeq
where $\cW(\tau,R,F)$ is some function to be determined. It automatically satisfies the diffeomorphism 
constraint. The Wheeler-DeWitt equation is second order in time derivatives so both positive and negative 
energy solutions exist, but we will confine our attention to the positive energy solutions above. It is 
worth noting that any functional
\beq
\Psi[\tau,R,\Gamma] = U\left(-\frac i 2 \int dr~ \Gamma(r)~ \cW(\tau,R,F)\right)
\label{genform}
\eeq
would satisfy the diffeomorphism constraint provided that $\cW$ has no explicit dependence on 
the label coordinate $r$ except through the mass function, $F(r)$. We have chosen $U=\exp$
so that the wave-functional may also be factorizable on a spatial lattice, whose cell size we
call $\sigma$, taking $\sigma\rightarrow 0$ in the continuum limit. Diffeomorphism invariance 
requires that the continuum wave-functional and all physical results be independent of the cell 
size. On the lattice, the argument of the exponential function becomes \cite{vws04,kmv06}
\beq
\int dr~ \Gamma(r)~ \cW(\tau,R,F) \rightarrow \sigma \sum_j \Gamma_j \cW(\tau_j,R_j,F_j),
\eeq
where $\Gamma_j = \Gamma(r_j)$, etc. This turns the wave-functional into a product state,
\beq
\Psi[\tau,R,\Gamma] = \prod_j \psi_j(\tau_j,R_j,F_j) = \prod_j \psi^{(0)}_j \exp\left[-\frac i2 \sigma 
\sum_j \Gamma_j \cW(\tau_j,R_j,F_j)\right]
\label{solnform}
\eeq
provided that $U=\exp$.

Before proceeding further it is necessary to define what is meant by a functional derivative when 
functions are defined on a lattice \cite{vws04}. The defining equations can be understood by analogy with the 
simplest properties of functional derivatives of the functions $J(x)$
\bea
&&\frac{\delta J(y)}{\delta J(x)} = \delta(y-x),\cr
&&\frac{\delta}{\delta J(x)}\int dyJ(y)  = 1
\eea
and from these definitions follows
\beq
\frac{\delta}{\delta J(x)}\int dyJ(y)\phi(y)=\phi(x).
\eeq
On a lattice we define, for the lattice intervals $x_i$ and $x_j$,
\beq
\frac{\delta J(x_{i})}{\delta J(x_{j})}=\Delta(x_i-x_j) = \lim_{\sigma \rightarrow 0}
\frac{\delta_{ij}}\sigma,
\label{funder}
\eeq
where $r_i$ labels the $i^\text{th}$ lattice site and $\delta_{ij}$ is the Kronecker $\delta$, equal 
to zero when the lattice sites $x_i$ and $x_j$ are different and one when they are the same. 
Just as $\delta(y-x)$ is only defined as an integrand in an integral, so $\Delta(x_{i}-x_{j})$ 
should also be considered defined only as a summand in a sum over lattice sites. Hence
\beq
\lim_{\sigma\rightarrow 0} \frac{\delta}{\delta J(r_j)} \sigma\sum_i J(r_i) = 
\lim_{\sigma \rightarrow 0}\sigma\sum_i \frac{\delta J(r_i)}{\delta J(r_j)}=1
\eeq
and
\beq
\frac{\delta}{\delta J(r_j)} \sigma\sum_i  J(r_i)\phi(r_i) = \lim_{\sigma\rightarrow 0} \sigma \sum_i 
\Delta(r_i-r_j)\phi(r_i) = \phi(r_j).
\eeq
It follows that
\beq
\frac{\delta}{\delta J(x_{j})}\rightarrow \frac 1\sigma\lim_{\sigma \rightarrow 0} \frac{\partial}
{\partial J_j},
\label{funder2}
\eeq
where $J_j=J(x_j)$. This is compatible with the formal (continuum) definition of the
functional derivative. 

\subsection{Collapse Wave Functionals}

When \eqref{funder} and \eqref{funder2} are applied to the Wheeler-DeWitt equation in \eqref{WDW_eq}
and $\Psi[\tau,R,\Gamma]$ is taken to be a product state, one obtains an equation describing the wave 
functions at each lattice point
\beq
\left[\frac{\partial^2}{\partial\tau_j^2} + \cF_j \frac{\partial^2}{\partial R_j^2} + A_j 
\frac{\partial}{\partial R_j} + B _j + \frac{\sigma^2\Gamma_j^2}{4\cF_j}\right]\psi_j \approx 0,
\label{wd1}
\eeq
but there is a further restriction arising from the diffeomorphism constraint. Inserting the ansatz 
in \eqref{solnform} into \eqref{wd1}, we find
\bea
&&\frac{\sigma^2\Gamma_j^2}4 \left[\left(\frac{\partial \cW_j}{\partial \tau_j}\right)^2 + \cF_j
\left(\frac{\partial \cW_j}{\partial R_j}\right)^2 - \frac 1\cF\right]\cr
&&\hskip 2cm \frac{\sigma\Gamma_j}2\left[\frac{\partial^2 \cW_j}{\partial \tau_j^2} + 
\cF_j\frac{\partial^2 \cW_j}{\partial R_j^2} + A_j \frac{\partial \cW_j}{\partial R_j}\right] + B_j = 0,
\eea
which must be satisfied {\it independently} of $\sigma$. This is only possible if the following 
three equations are simultaneously satisfied at each lattice site \cite{kmv06},
\bea
&&\left[\left(\frac{\partial \cW_j}{\partial\tau_j}\right)^2 + \cF_j 
\left(\frac{\partial \cW_j}{\partial R_j}\right)^2 - \frac{1}{\cF_j}\right]=0,\cr
&&\left[\frac{\partial^2 \cW_j}{\partial\tau_j^2} + \cF_j\frac{\partial^2 \cW_j}
{\partial R_j^2} + A_j \frac{\partial \cW_j}{\partial R_j} \right]  =  0,\cr
&& B_j = 0.
\label{4eqns}
\eea
Moreover, it is straightforward that the Hamiltonian constraint is Hermitean if and only if 
\beq
A_j = \cF_j\partial_{R_j} \ln ({\mathfrak m}_j |\cF_j|),
\label{meas}
\eeq
where $\mathfrak m_j$ is the Hilbert space measure.

Unique solutions to the equations in \eqref{4eqns} and having the form given in \eqref{solnform} 
have been obtained in all, even the non-marginally bound, cases \cite{kmv06}. For the marginally 
bound models the solution for the phase $\cW_j$ in the exterior, {\it i.e.,} for shells that lie 
outside the apparent horizon ($R_j>F_j$), is
\beq
\cW_j^{(\pm)} = \tau_j \pm 2F_j\left[z_j -\tanh^{-1}\frac 1{z_j}\right],~~ z_j > 1,
\label{out}
\eeq
where $z_j=\sqrt{R_j/F_j}$. The positive sign refers to ingoing waves, traveling toward the horizon and 
the negative sign to outgoing waves, as can be seen from the signature of the phase velocity,
\beq
\dot z_j = \mp \frac{z_j^2-1}{2F_jz_j^2},
\eeq
keeping in mind that $z_j>1$. The phase velocity approaches zero at the horizon. In the interior,  
{\it i.e.,} for shells that lie inside the apparent horizon  ($R_j<F_j$), the solution is 
\beq
\cW_j^{(\pm)} = \tau_j \pm 2F_j\left[z_j -\tanh^{-1} z_j \right],~~ z_j < 1,
\label{in}
\eeq
but here the the positive sign refers to outgoing waves and the negative sign to ingoing waves, traveling 
toward the central singularity, again as determined by the phase velocity. Furthermore, 
as shown in Appendix B of \cite{kmv06},  the system in \eqref{4eqns} determines not only  $\cW_j$ but 
the Hilbert space measure, $\mathfrak{m}_j$, as well. For the marginal models under consideration, 
$\mathfrak{m_j}$ is regular everywhere and given by
\beq
\mathfrak m_j = z_j
\label{meas}
\eeq
up to a constant scaling.

\section{Hawking Radiation}

In this section we will argue that the states described above yield Hawking radiation. Our 
first approach will closely parallel Hawking's original work \cite{haw75}. First we need to introduce the 
concept of a black hole into the formalism above, which, following \cite{vksw03}, we do by taking the 
mass function to be of the form
\beq
F(r) = 2 M \Theta(r)+f(r),
\label{mf0}
\eeq
where $\Theta(r)$ is the Heaviside function and $f(r)$ represents a dust perturbation ($f(r)/2M \ll 1$).
It is easy to see that with this choice of mass function, the black hole state factors out in 
\eqref{solnform} and the remaining state then assumes the same form with $F$ replaced by $2M$ and 
$\Gamma$ replaced by $f'(r)$,
\beq
\psi[\tau,R,\Gamma] = e^{\pm i M \cW_0}\times \exp\left[\pm \frac i2 \int dr f'(r) W^f(\tau,R,M)\right]~~
\stackrel{\text{def}}{=}~~ \Psi_\text{bh} \times \Psi_f,
\eeq
where $\cW_0=\cW(\tau(0),R(0),F(0))$ and the first exponent represents the black hole at the 
origin and the second, up to order $f(r)$, represents the matter distribution that propagates in this background if 
we take $F(r) \approx 2M$ in $\cW^f$.

Next, we must identify those quantum states that correspond to the ingoing and outgoing modes, respectively
and evaluate an appropriate inner product. Since the description should refer to observers at infinity, the 
inner product will be evaluated on hypersurfaces of constant Killing (Schwarzschild) time $T$, and not on 
hypersurfaces of constant dust time (which corresponds to freely falling observers). It is not difficult to 
show that for contracting clouds the an infalling wave is given by
\beq
\Psi_f = \exp\left[-i \int dr f'(r)\left(T + 8M\left(z-\tanh^{-1}\frac 1z\right)\right)\right]
\eeq
being approximately 
\beq
\Psi_f^- \approx \exp\left[-i \int dr f'(r)\left(T + 8Mz\right)\right]
\label{ass}
\eeq
when $T\rightarrow -\infty$ and $z\rightarrow \infty$ and 
\beq
\Psi_f^+ \approx \exp\left[-i \int dr f'(r)\left(T - 8M\tanh^{-1}\frac 1z\right)\right]
\label{asp}
\eeq
when $T\rightarrow +\infty$ and $z\rightarrow 1$. Thus, the simple looking phase on $\Im^-$ scatters 
through the geometry to turn into the complicated looking phase on $\Im^+$ near the horizon. This is 
similar to what happens in the geometric optics approximation.

Equation (\ref{ass}) represents infalling waves. We can think of it as a product over plane waves, 
one at each label $r$, as follows:
\beq
\Psi_\omega^{-}=\prod_j e^{-i\omega_j [T_j+8Mz_j]}.
\label{pw}
\eeq
This should represent a complete set of infalling modes at each label $j$, if we think of
the $\omega_j$ as the frequency of the modes. A complete set of outgoing modes on $\Im^+$ 
would likewise be given by the functional
\beq
{\Psi}_\omega^{+}=\prod_j e^{-i\omega_j[T_j-8Mz_j]}\ .
\label{og}
\eeq
We now ask: what is the projection of our solution \eqref{asp} on the negative frequency modes of 
the outgoing basis on $\Im^+$. For this purpose we must consider the inner product of states on a 
hypersurface of constant Schwarzschild time $T$. It turns out that Hawking's thermal radiation is 
recovered if take the metric in the $(\tau,R)$ plane is given by the quadratic term in the 
Hamiltonian constraint \eqref{H_new} instead of \eqref{meas}. We speculate that this has to do with 
the essentially classical role played by the black hole, described by $\Psi_\text{bh}$. Transforming 
to the metric in $(R,T)$ coordinates we find
$$ g_{RR} = \left({R\over R-2M}\right)^{2}.$$
The required Bogoliubov coefficient is then given by the following inner product on a constant 
$T$ hypersurface,
\beq
\beta(f,\omega) = \left\langle \Psi_\omega\vert\Psi _f^{+}\right\rangle =\prod_j \int \sqrt{g_{RR}}~dR_j ~
\Psi_{\omega_j} ^{+}\Psi_{f_j}^{+},
\eeq
which represents the negative frequency modes present in \eqref{asp}. A straightforward computation
then yields
\beq
\langle \text{in}|\widehat N_\text{out}|\text{in}\rangle = |\beta(f,\omega)|^2 \approx \prod_j 
\frac{2\pi M}{\Delta f_j}\left[\frac 1{e^{8\pi M \Delta f_j}-1}\right],
\eeq
which is interpreted as the eternal black hole being in equilibrium with a thermal bath at the Hawking 
temperature $(8\pi M)^{-1}$. Thus we have a functional Schr\"odinger picture for dust Hawking radiation
(see \cite{kmsv07} for a generalization to the non-marginal models).

An alternative approach, one that is better adapted to quantum {\it collapse}, was considered 
in \cite{vw10}. By matching the shell wave functions describing 
gravitational collapse across the apparent horizon, it was shown that an ingoing wave on one side of 
the apparent horizon is necessarily accompanied by an outgoing wave on the other side. Furthermore,
the relative amplitude of the outgoing wave is suppressed by the square root of the Boltzmann factor
at the ``Hawking'' temperature of the shell. Strictly speaking the Hawking temperature, $T_H=(8\pi GM)^{-1}$,
refers to an eternal black hole of mass $M$. The temperature appearing in the Boltzmann factor from
matching shell wave functions across the horizon is $T_H = (4\pi G F)^{-1}$, where $F$ is the mass
function and represents twice the mass contained within the shell. Diffeomorphism invariant wave 
functionals describing the collapse can also be matched and yields the same picture, but now the 
relative amplitude of the outgoing wave functional to the ingoing one is given by $e^{-S/2}$, where $S$ 
is the entropy of the final state black hole.

\section{Black Hole Entropy}

As mentioned in the previous section, black holes with ADM mass parameter $M$ are special cases of 
the solution in \eqref{ltb-metric}, obtained when the mass function is constant, $F=2~GM$, and the 
energy function is vanishing. This can be shown directly by a coordinate transformation of 
\eqref{ltb-metric} from the comoving system $(\tau,\rho)$ to static coordinates $(T,R)$, in which the 
metric has the standard from,
\beq
ds^2 = \cF(R)dT^2 - \cF^{-1}(R) dR^2 - R^2 d\Omega^2.
\label{bh2}
\eeq
We imagine therefore that the eternal black hole is a single shell and represented by the mass 
function
\beq
F(r) = 2M \Theta(r)
\label{massfn2}
\eeq
($G=1)$, where $M$ is the mass at label $r=0$ and $\Theta(r)$ is the Heaviside function. The mass density function 
is therefore 
\beq
\Gamma(r) = 2M \delta(r)
\eeq
and, because of the $\delta-$distributional mass density, the wave-functional in \eqref{ansatz} turns
into the wave-{\it function}
\beq
\Psi[\tau,R,\Gamma] = e^{-\frac i2\int_0^\infty dr \Gamma(r) \cW(\tau(r),R(r),F(r))}
 = e^{-i M\cW_0(\tau,R,F)}
\eeq
where $\tau=\tau(0)$, $R=R(0)$ and $F=F(0)$. The Wheeler-DeWitt equation now becomes the Klein-Gordon
equation describing the shell. Taking into account the factor ordering ambiguities and absorbing 
the $M$ dependent term, which now renormalizes the potential, into the function $B(R,F)$ we have
\beq
\left[\frac{\partial^2}{\partial \tau^2}+\cF\frac{\partial^2}{\partial R^2} + A
\frac{\partial}{\partial R}+B\right]e^{-i M\cW_0(\tau,R,F)} = 0,
\label{wd1}
\eeq
In contrast with the case in which the mass density is a smooth function over some set of non-zero 
measure, no regularization is necessary here. This means that no further conditions must be met
and therefore that the measure as well as the functions $A(R,F)$ and $B(R,F)$ will remain undetermined 
although the function $A(R,F)$ will continue to be related to the measure according to \eqref{meas}.
Thus two conditions are required to proceed with the quantization of the black holes as described above.
 
The first condition we impose is one on the measure appropriate to the Hilbert space of wave-functions.
In the previous section we obtained the Hawking evaporation of a collapsing dust cloud surrounding a pre-existing 
black hole by taking the dust as a small perturbation to the black hole mass function in \eqref{massfn2}. 
The calculation proceeded by evaluating the Bogoliubov coefficient in the near horizon limit outside 
the horizon and crucial to obtaining the correct Hawking temperature is the choice of measure appropriate 
for eternal black holes. The measure was obtained from the DeWitt supermetric, $\gamma_{ab}$, on the 
configuration space $(\tau,R)$ and can be read directly from the Hamiltonian constraint
\beq
\gamma_{ab} = \left(\begin{matrix}
1 & 0 \cr
0 & 1/\cF\end{matrix}\right).
\eeq
It gives $\mu = 1/\sqrt{|\cF|}$, {\it i.e.,}
\beq
\langle \Psi_1,\Psi_2\rangle = \int \frac{dR}{\sqrt{|\cF|}} \Psi_1^\dagger \Psi_2
\eeq
as well as the function $A(R,F)$ via the hermiticity condition \eqref{meas}.
As long as $\cF\neq 0$ the Wheeler DeWitt equation can now be written as 
\beq
\left[\frac{\partial^2}{\partial \tau^2} \pm \frac{\partial^2}{\partial R_*^2} + B\right] \Psi = 0,
\eeq
where the positive sign in the above equation refers to the exterior, while the negative 
sign refers to the interior and $R_*$ is defined by
\beq
R_* = \pm \int \frac{dR}{\sqrt{|\cF|}}.
\eeq
The second condition arises because we are describing a single shell in this simple quantum 
mechanical model of an eternal black hole and because $B(R,F)$ represents an interaction of 
the shell with itself. We simply demand there are no self interactions, {\it i.e.,} that 
$B(R,F)=0$. The quantum evolution is then described by the free wave equation in the interior, 
but by an elliptic equation in the exterior. This signature change has been noted in other 
models \cite{ki89,bk97} and occurs because of the behavior of $\cF$, which passes from positive 
outside the horizon to negative inside. For the black hole, it means that its wave function 
is supported in its interior. The spectrum will be determined by the proper radius, $L_h$, 
of the horizon,
\beq
L_h(M) = \int_0^{R_h} \frac{dR}{\sqrt{|\cF|}},
\label{propl}
\eeq
where $R_h$ is its area radius. If we extend the coordinate $R_*$ to range over $(-\infty,\infty)$, 
thereby avoiding any issues related to a boundary at the center, this simple model of a 
quantum black hole effectively describes a dust shell in a ``box'' of radius $2L_h(M)$, which itself 
depends on its total ADM mass. The stationary states describe a spectrum of the form (reintroducing 
$\hbar$ and $G$)
\beq
4G M_j L_{h,j} = A_\text{Pl} \left(j+\frac 12\right),
\label{quant1}
\eeq
where $j$ is a whole number and $A_\text{Pl}=h G$ is the Planck area. It is straightforward to show 
that $L_h = \pi G M$, so
that
\beq
\frac{A_j}4 = A_\text{Pl}\left(j+\frac 12\right),
\label{areaquant}
\eeq
where $A_j$ is the horizon area \cite{vw01}. 

As we show below, the equispaced area spectrum predicted by our simplified model of a quantum black hole 
implies that the entropy obeys the Bekenstein-Hawking area law provided that the area quanta are assumed 
distinguishable. The entropy therefore also admits a discrete and equispaced spectrum. What is the origin 
of the degeneracy that leads to the black hole entropy?
Although the black hole is treated as a single shell with the spectrum in \eqref{quant1}, this single shell 
is in fact the end state of many shells that have collapsed to form the black hole. Regardless of their 
history, we assume that each of the shells then occupies only the levels of \eqref{quant1}, contributing 
some multiple of the Planck area to the total horizon area of the final 
state. A black hole microstate is thus a particular distribution of collapsed shells among the available 
levels. If the distribution of shells is such that $\cN_j$ shells occupy level $j$, the black hole's 
total horizon area becomes 
\beq
\frac A4 = A_\text{Pl} \sum_j \left(j+\frac 12\right) \cN_j
\label{totA}
\eeq
and the (single shell) solution in \eqref{bh2} is to be interpreted as an excitation by $\cN=\sum_j \cN_j$ 
collapsed shells. 

The spectrum in \eqref{totA} represents the ``area ensemble'' and the number of black hole microstates 
giving the ``area'' $A$ will depend on assumptions concerning the degeneracy of the microstates.  Assuming 
the shells to be distinguishable, the number of states can be written in terms of the total number of area 
quanta, $\cQ$, and the total number of shells, $\cN$, as
\beq
\Omega(\cQ,\cN) = \frac{(\cN+\cQ-1)!}{(\cN-1)!\cQ!},
\eeq
where 
\beq
\cQ=\frac{A}{A_\text{Pl}} - \frac{\cN}2.
\eeq
Holding $A$ fixed and extremizing the microcanonical entropy, $S_\text{micro}=k_B \ln \Omega$, 
with respect to the number of shells gives
\beq
S_\text{micro} = (2k_B \coth^{-1}\sqrt{5})\frac A{4 A_\text{Pl}},
\label{entmicro}
\eeq
which is in excellent (better than 96\%) agreement with the Bekenstein-Hawking 
entropy. In addition to the exponential growth in the number of states, the area quantization in 
\eqref{totA} ensures that the entropy is effectively quantized in units of the Planck area, as 
originally proposed by Bekenstein \cite{bek73}. 

Note that it is quite a simple matter to show that had the shells been assumed indistinguishable 
then the entropy would depend on the square-root of the area \cite{Kolland}. The fact that the 
area degrees of freedom must be treated as distinguishable runs contrary to our intuition for elementary 
degrees of freedom in quantum field theory and calls into question whether ``area'' is a fundamental 
quantity in quantum gravity.

\section{Discussion}

In our contribution we have described a canonical quantization of collapsing, inhomogeneous dust
and some of its implications. While here we have confined ourselves to four dimensions and a vanishing 
cosmological constant, our set up can be extended to describe dust collapse in any dimension
both with and without a cosmological constant and in all cases it is possible to show that static 
black holes will radiate at the appropriate Hawking temperature of the black hole \cite{vts08,fgk10},
at least in the semi-classical approximation in which the mass function is taken to be of the form given 
in \eqref{mf0}. Two, likely related, features of this description remain to be understood. Firstly, 
why does the Hawking picture rely on the measure derived from the classical Hamiltonian constraint to 
describe eternal black holes? Secondly how does the semiclassical picture hold up {\it during} collapse, 
{\it i.e.,} when the 
dust is not taken to be a perturbation on the background of a massive black hole? In principle, the 
second question could be addressed by pursuing the approach of \cite{vw10}, but it still remains to be done.
Another related problem concerns the regularization used to define the functional Wheeler-DeWitt equation
\eqref{WDW_eq}. The lattice regularization we have used is an {\it ad hoc} regularization in which the 
divergent terms have to cancel each other. It leads directly to \eqref{4eqns}, implying that the Hamilton 
Jacobi solution is exact and, if factorizability on the lattice is required, making \eqref{out} and 
\eqref{in} unique as well. However, regularization independence of the results described here remains an
open question and factorizability on a spatial lattice may be too strong a condition to impose.

Fundamental questions concerning the nature of the quantum gravitational degrees freedom remain, and 
these are most manifest in the assumptions that must be made concerning the statistics obeyed by them. 
For the BTZ black hole in 2+1 dimensions it is necessary to employ Bose statistics 
instead of Boltzman statistics in order to recover the entropy \cite{vgksw08}. This corresponds to the fact 
that the BTZ black hole admits an equispaced {\it mass} spectrum (as opposed to an area spectrum) and 
must be treated in the energy ensemble. Its heat capacity is positive, whereas it is negative for the 
Schwarzschild black hole. The actual computation of the entropy is similar to that employing the AdS/CFT 
correspondence \cite{s98} and the entropy is found to depend on two quantities, {\it viz.,} the energy of 
what is taken to be the vacuum solution, $\Delta_0$, and a constant, $M_0$, arising out of a boundary 
contribution at the origin.\footnote{Whereas a boundary contribution from the origin in 3+1 dimensional 
collapse would represent a singular initial configuration and is therefore set to zero, in 2+1 dimensions 
a non-vanishing contribution is essential to allow for an initial velocity profile that vanishes there. 
This does not lead to singular initial data and the boundary contribution does not have the interpretation 
of a point mass situated at the center \cite{g05}.} A comparison between the result from the canonical theory
and the result obtained via the AdS/CFT correspondence yields an effective central charge 
\beq
c_\text{eff} = \frac 12\left[1-\frac{48l}{\hbar}(M_0-\Delta_0)\right],
\eeq
which must be set to $3l/2G\hbar$ to achieve agreement with the Bekenstein-Hawking entropy. This is the 
central charge of the Liouville theory induced at spatial infinity by 2+1 dimensional gravity 
\cite{chd95}

In three or more spatial dimensions, with a negative cosmological constant, the spectrum of states
describing eternal black holes has a more complicated description \cite{vw09}. For small black holes, 
by which we mean black holes whose horizon radii are much smaller than the AdS length, 
an approximate area spectrum, similar to \eqref{areaquant}, is obtained and the entropy is recovered by 
working in an area ensemble in which the degrees of freedom are assumed distinguishable. In this case, 
the black hole heat capacity is negative. On the other hand, for large AdS black holes, {\it i.e.,} 
black holes whose horizon radii are much larger than the AdS length, it is the {\it mass} that is 
quantized in integer units and the (mass) spectrum turns out to be independent of the gravitational 
constant, $G$. This is similar to what happens with the BTZ black hole (for which, however, there is 
no ``small black hole'' limit). The entropy is now obtained in an energy ensemble and the degrees of 
freedom must be assumed indistinguishable. It can then be shown that the thermodynamics in the large black 
hole limit is inextricably connected with the thermodynamics in the opposite limit by a duality of the 
Bose partition function. The gravitational constant, $G$, absent 
in the mass spectrum, reemerges in the thermodynamic description via this duality and the black hole 
heat capacity is positive. It appears that the Hawking-Page transition \cite{hawpag83} separates 
the particle-like degrees of freedom of large black holes, which must be counted as indistinguishable, 
from the geometric degrees of freedom of small black holes, which are counted as distinguishable.
This remains rather mysterious and seems worth pursuing further, but that is complicated by the fact 
that it is difficult to  obtain closed form solutions for the spectrum of AdS black holes between the 
two limits described.

In conclusion, we now seem to be in a position to address some of the issues that have surrounded Black 
Hole thermodynamics for decades, such as the information loss problem, questions about singularity avoidance 
and the radiation from a naked singularity, at least within the context of a special class of models. 
We expect progress in addressing these and the new issues that have arisen since developing the quantization 
described here to be made in the near future. 


\bigskip

\begin{thebibliography}{99}

\bibitem{l33}G. LeMa\^\i tre, Annales de la Soci´et´e Scientifique de Bruxelles A 53, 51 (1933); for an English
translation see Gen. Rel. Grav. 29, 641 (1997).
\bibitem{k97}A. Krasinski, {\it Inhomogeneous Cosmological Models} (Cambridge University Press, Cambridge,
1997).
\bibitem{k95}K. V. Kucha\v r, Phys. Rev. D {\bf 50} (1994) 3961.
\bibitem{r95}J.D. Romano, \href{http://arxiv.org/abs/gr-qc/9501015v1}{[arXiv:gr-qc/9501015]}
\bibitem{js95}P. Joshi and T.P. Singh, Phys. Rev. {\bf 51} (1995) 6778 .
\bibitem{vws01}C. Vaz, L. Witten and T.P. Singh, Phys. Rev. D {\bf 63} (2001) 104020.
\bibitem{kmv06}C. Kiefer, J. M-Hill and C. Vaz, Phys. Rev. D {\bf 73} (2006) 044025.
\bibitem{bk95}J.D. Brown and K.V. Kucha\v r, Phys. Rev. D {\bf 51} (1995) 5600.
\bibitem{OUP}C. Kiefer, {\it Quantum Gravity} (Clarendon Press, Oxford, 2004).
\bibitem{tw87}N. C. Tsamis and R. P. Woodard, Phys. Rev. D {\bf 36} (1987) 3641.
\bibitem{vws04}C. Vaz, L. Witten and T.P. Singh, Phys. Rev. D {\bf 69} (2004) 104029.
\bibitem{haw75}S. W. Hawking, Comm. Math. Phys. {\bf 43} (1975) 199.
\bibitem{vksw03}C. Vaz, C. Kiefer, T.P. Singh and L. Witten, Phys. Rev. D {\bf 67} (2003) 024014.
\bibitem{kmsv07}C. Kiefer, J. Mueller-Hill, T.P. Singh and C. Vaz, Phys. Rev. D {\bf 75} (2007) 124010.
\bibitem{vw10}C. Vaz and L.C.R. Wijewardhana, Phys. Rev. D {\bf 82} (2010) 084018.
\bibitem{ki89} C. Kiefer, Phys. Lett. B {\bf 225}, 227 (1989).
\bibitem{bk97}T. Brotz and C. Kiefer, Phys. Rev. D {\bf 55} (1997) 2186.
\bibitem{vw01}C. Vaz and L. Witten, Phys. Rev. D {\bf 64} (2001) 084005.
\bibitem{bek73}J. Bekenstein, Phys. Rev. D {\bf 7} (1973) 2333.
\bibitem{Kolland} C. Kiefer and G. Kolland, Gen. Rel. Grav. {\bf 40} (2008) 1327.
\bibitem{vts08}C. Vaz, R. Tibrewala and T.P. Singh, Phys. Rev. D {\bf 78} (2008) 024019.
\bibitem{fgk10}A. Franzen, S. Gutti and C. Kiefer, Class. Quant. Grav. {\bf 27} (2010) 015011.  
\bibitem{vgksw08}C. Vaz, S. Gutti, C. Kiefer, T.P. Singh and L.C.R. Wijewardhana, Phys. Rev. D {\bf 77}
(2008) 064021.
\bibitem{s98}A. Strominger, JHEP 9802 (1998) 009.
\bibitem{g05}S. Gutti, Class. Quant. Grav. {\bf 22} (2005) 3223.
\bibitem{chd95}O. Coussaert, M. Henneaux, and P. van Driel, Class. Quant. Grav. {\bf 12} (1995) 2961.
\bibitem{vw09}C. Vaz and L.C.R. Wijewardhana, Phys. Rev. D {\bf 79} (2009) 084014.
\bibitem{hawpag83}S. W. Hawking and D. N. Page, Commun. Math. Phys. {\bf 87} (1983) 577.
\end{thebibliography}
\end{document}